\documentclass[12pt,preprint]{aastex}
\usepackage{amssymb}

\begin{document}
\title{Density Estimation for Projected Exoplanet Quantities}

\author{Robert A.\ Brown}
\affil{Space Telescope Science Institute, 3700 San Martin Drive, Baltimore, MD 21218}
\email{rbrown@stsci.edu}

\begin{abstract}
Exoplanet searches using radial velocity (RV) and microlensing (ML) produce samples of ``projected'' mass and orbital radius, respectively. We present a  new method for estimating the probability density distribution (density) of the unprojected quantity from such samples. For a sample of $n$ data values, the method involves solving $n$ simultaneous linear equations to determine the weights of delta functions for the raw, unsmoothed density of the unprojected quantity that cause the associated cumulative distribution function (CDF) of the projected quantity to exactly reproduce the empirical CDF of the sample at the locations of the $n$ data values.  We smooth the raw density using nonparametric kernel density estimation with a normal kernel of bandwidth $\sigma$. We calibrate the dependence of $\sigma$ on $n$ by Monte Carlo experiments performed on samples drawn from a theoretical density, in which the integrated square error is minimized. We scale this calibration to the ranges of real RV samples using the Normal Reference Rule. The resolution and amplitude accuracy of the estimated density improve with~$n$. For typical RV and ML samples, we expect the fractional noise at the PDF peak to be approximately 80 $n^{-\log 2}$. For illustrations, we apply the new method to 67 RV values given a similar treatment by Jorissen et~al.\ in 2001, and to the 308 RV values listed at exoplanets.org on 20 October 2010.  In addition to analyzing observational results, our methods can be used to develop measurement requirements---particularly on the minimum sample size $n$---for future programs, such as the microlensing survey of Earth-like exoplanets recommended by the Astro 2010 committee. 
\end{abstract}

\keywords{astrobiology --
methods: data analysis --
methods: statistical --
techniques: radial velocity --
(stars:) planetary systems --
(stars:) binaries: spectroscopic}

\section{INTRODUCTION}


Mass $m$ and orbital radius $r$ are two key factors for the habitability of 
exoplanets. This is because $m$ plays an important role in the retention of an atmosphere, and $r$ is a key determinant of the surface temperature. Besides those connections to the cosmic search for life, the true distributions of $m$ and $r$ are also important for theories of the formation and dynamical evolution of planetary systems. Therefore, we have a variety of good reasons to better understand the cosmic distributions of $m$ and $r$. Such improvement will involve learning more from measurements already made, as well as anticipating results from the  telescopes and observing programs of the future. 

We focus on two astronomical techniques that measure \textit{projected} planetary quantities. One is radial velocity (RV), the source of many exoplanet discoveries so far. The other is microlensing (ML). Astro 2010 recently recommended a ML survey for Earth-mass exoplanets on orbits wider than detectable by \textit{Kepler} (Blandford et~al.\ 2010). 

RV yields the unprojected value of $r$, from the observed orbital period and an estimate of the stellar mass, but for mass it can only provide $m\sin i$---not $m$---where $i$ is the usually unknown orbital inclination angle. Correspondingly, if the mass of the lens and the stellar distances are known or assumed, ML yields $m$, but for $r$ it usually can provide only $r\sin\beta$---not $r$---where $\beta$ is the usually unknown planetocentric angle between the star and observer (Gaudi 2011). For the foreseeable future, only an ML survey appears to offer observational access to exoplanets like Earth in mass and orbital radius around Sun-like stars. A high-confidence estimate of the occurrence probability of such planets around nearby stars is critical to designing future telescopes to obtain their spectra and search for signs of life.

Even when the projection angle $\beta$ is unknown, we can use statistical methods to draw inferences from samples of the projected values about the distribution of the true, unprojected values in nature. This paper presents a new method to help draw those inferences.

One goal of this paper is to introduce a science metric for the ML survey, and to demonstrate its use with existing RV results. The ML metric is the resolution and accuracy of the estimated probability density of the orbital radii of Earth-mass exoplanets. Such a science metric will be useful for setting measurement requirements, designing telescopes and instruments, planning science operations, and arriving at realistic expectations for the new ML survey program. 

\section{STARTING POINT}



Our starting point is recognizing that the projected quantity, $\varphi \equiv 
m\sin i$ or $r\sin \beta$, is the product of two independent  continuous random variables, $\rho \equiv m$ or $r$, and $y\equiv \sin i$ or $\sin\beta$, which have densities $\Psi (\rho)$ and ${\cal Q}(y)$, respectively. Assuming the directions of planetary radius vectors and orbital poles are uniformly distributed on the sphere, 
\begin{equation}
\label{eq1}
{\cal Q}(y)=\frac{y}{\sqrt {1-y^2} }~~.
\end{equation}
$\rho $ and $y$ have ranges $0\le \rho \le \infty$ and $0\le y\le 1$. The product $\varphi =\rho y$ is also a random variable, with the density $\Phi (\varphi)$, which we can calculate as follows. The probability density at a point $\{ {\rho, y} \}$ on the $\rho$--$y$ plane---within the ranges of $\rho$ and $y$---is $\Psi (\rho)\;{\cal Q}(y)$, and $\Phi (\varphi )$ is the 
integral of this product over the portion of the $\rho$--$y$ plane where $\rho y=\varphi $:
\begin{eqnarray}
 \label{eq2}
\Phi (\varphi )&=&\int_0^\infty {\int_0^1 {\Psi (\rho ){\cal 
Q}(y)} } \delta _\varphi (\rho y-\varphi )dy\, d\rho \nonumber\\
 &=&\int_0^\infty {\Psi (\rho)\int_0^1 {{\cal Q}(y)}} 
\delta _y \left( {y-\frac{\varphi }{\rho}} \right)\frac{1}{\rho}dy\, d\rho\\ 
&=&\int_\varphi ^\infty {\Psi (\rho){\cal Q}
\left( {\frac{\varphi}{\rho}} \right)} \frac{1}{\rho }\, d\rho~~,\nonumber 
 \end{eqnarray}
where {$\delta_{z}$ is the Dirac delta function for any variable $z$, with the 
normalization
\begin{equation}
\label{eq3}
\int_{-\infty }^\infty {\delta _z (z-a)dz=1}~~,
\end{equation}
and where in the last line of Eq.~\ref{eq2} we have used the fact that $\rho $ must always be greater than $\varphi$. We now change the variable $\varphi \to u\equiv \log \varphi$ with the density ${\cal P}(u)$:
\begin{eqnarray}
\label{eq4}
{\cal P}(u)&=&\Phi (\varphi)\frac{d\varphi}{du}=\Phi ({10^u})
(\ln 10)10^u \nonumber\\ 
&=&({\ln 10})10^u\int_{10^u}^\infty {\Psi (\rho){\cal Q}
\left({\frac{10^u}{\rho}}\right)} \frac{1}{\rho }\, d\rho~~. 
 \end{eqnarray}
We now change the variable $\rho \to x\equiv \log \rho$, with the density ${\cal R}(x)$. Using the relation $\Psi (\rho)\, d\rho ={\cal R}(x)\;dx$, we 
have
\begin{equation}
\label{eq5}
{\cal P}(u)={\cal C}(u)({\ln 10})\int_u^\infty {{\cal R}(x)
{\cal Q}({10^{u-x}})} 10^{u-x}dx~~.
\end{equation}
In Eq.~\ref{eq5}, we have introduced a new factor, the completeness function ${\cal C}(u)$, to account for variations in search completeness due to, for example, the variation of instrumental sensitivity with $u$. The clearest example is declining signal-to-noise ratio with smaller $u$---smaller $m \sin i$ in the case of RV and smaller $r \sin \beta$ for ML. In this paper, we ignore completeness effects and assume ${\cal C}(u)\equiv 1.$ In the case of the density estimation studies in Section~4, this means we have not necessarily chosen a \textit{realistic} theoretical ${\cal R}(x)$, but realism is probably not necessary for the immediate purposes---achieving a valid calibration of the bandwidth (smoothing length) $\sigma$ and exploring resolution and accuracy. In the case of the analysis of real RV data in Section~5, we need only to remember that scientific interpretations of the distributions we infer for ${\cal R}(\log m)$ will demand qualification regarding the potential effects of ${\cal C}(u)$, particularly for the left-hand tails of the distribution, where incompleteness due to low signal-to-noise ratio must be important.

We recognized that Eq.~(\ref{eq5}) has a thought-provoking analogy to the case of an astronomical image (which corresponds to ${\cal P}$), where the object (${\cal R})$ is convolved with the telescope's point-spread function (${\cal Q}$) and the field is, say, vignetted in the camera (${\cal C}$). Pursuing this analogy, we might call the form of the integral in Eq.~(\ref{eq2})  ``logarithmic convolution.'' It could be said that we ``see'' the true distribution (${\cal R}$) only after it has been logarithmically convolved with the projection function (${\cal Q}$) and modulated by the completeness function (${\cal C}$). As with image processing, we can correct ``vignetting'' by dividing ${\cal P}$ by ${\cal C}$, if we know it, and then ``deconvolve'' the result to remove the effects of ${\cal Q}$---which in this case we know \textit{exactly}. As with image processing, the result can be a \textit{transformation}---a new, alternative, precise description of the sample, in the form of an estimate of the ``object,'' the natural density ${\cal R}$, with some systematic effects reduced or removed (but other effects possibly remaining).

Equation~(\ref{eq5}) is a form of Abel's integral equation, as discussed in this general context by Chandrasekhar \& M\"unch (1950), who used the formal solution to investigate the distribution of true and apparent rotational velocities of stars. Later, Jorissen et~al.\ (2001) inferred the distribution of exoplanet masses from 67 values of $m \sin i$, using both the formal solution and the Lucy-Richardson algorithm, which is an implementation of the expectation-maximization algorithm that is widely used in maximum-likelihood estimation (Dempster et~al.\ 1977). We present a third numerical approach to solving Eq.~(\ref{eq5}).

In Section 3, we develop the new method for transforming samples of measured values into a raw, unsmoothed density for $x$. In Section~4, we discuss nonparametric density estimation, random deviates, and the qualitative accuracy of the estimated density.  In Section~5, we illustrate these computations on samples of RV results---the 67 values treated by Jorissen et~al.\ (2001) and the 308 values of $m\sin i$ available at exoplanets.org on 20~October 2010. 
In Section~6, we comment on implications and future directions.

\section{NEW METHODS}

We assume the sample of independent and identically distributed, projected values $\{u\}$ is non-redundant and 
sorted in ascending order. The cardinality of the sample is $n$. The empirical cumulative 
distribution function (CDF) for $u$ is
\begin{equation}
\label{eq6}
\hat{P}(u)\equiv \frac{1}{n}\sum\nolimits_{i=1}^n {\cal I}(u_i \le u)~~,
\end{equation}
where ${\cal I}$ is the indicator function
\begin{equation}
\label{eq7}
{\cal I}(\mathit{logical\ statement})\equiv 1\ \mathrm{if\ the\ statement\ is\ true\ and\ 0\ otherwise}.
\end{equation}
$\hat{P}$ estimates the true CDF of the projected quantity, which is
\begin{equation}
\label{eq8}
P(u)\equiv \int_{-\infty }^u {\cal P}(u')du'~~.
\end{equation}

We can approximate ${\cal R}$ by a sum of Dirac delta functions at $N$ points $\{x\}$ with weights $\{w\}$:
\begin{equation}
\label{eq9}
{\cal R}(x)\simeq \hat{\cal R}(x)\equiv \sum\nolimits_{j=1}^N w_j \delta 
(x-x_j)~~.
\end{equation}
The associated approximation of $P(u)$ is
\begin{eqnarray}
\label{eq10}
P(u)\simeq \hat{P}'(u) &\equiv &\int_{-\infty }^u {\ln 10\int_{{u}'}^\infty 
{\sum\nolimits_{j=1}^N {w_j \delta (x-x_j)} {\cal Q}(10^{{u}'-x})} 
10^{{u}'-x}dxd{u}'} \\ 
&= &\sum\nolimits_{j=1}^N {w_j \ln 10\int_{-\infty }^u {\frac{10^{2({u}'-x_j )}}{\sqrt {1-10^{2({u}'-x_j )}} }} } {\cal H}(-{u}'+x_j )d{u}'\nonumber \\ 
\noalign{\smallskip}
&= &\sum\nolimits_{j=1}^N {w_j \left( {1-\sqrt{1-10^{2(u-x_j )}} {\cal 
H}(-u+x_j )} \right)}~~,\nonumber 
\end{eqnarray}
where 
\begin{equation}
\label{eq11}
{\cal H}(-a+b)\equiv 1\mbox{ if }a\le b\mbox{ and }0\mbox{ if }a>b
\end{equation}
is the Heaviside unit-step function. When we set $N=n$, $u=u_i$, and $x_j$, the critically determined set of linear equations
\begin{equation}
\label{eq12}
\sum\nolimits_{j=1}^N {w_j \left( {1-\sqrt {1-10^{2(u_i-u_j )}}  H(-u_i+u_j )} \right)} =\frac{1}{n}\sum\nolimits_{k=1}^n  I(u_k \le u_i)
\end{equation}
can be solved for $\{w\}$ by multiplying the vector that is the right side of Eq.~(\ref{eq12}) by the inverse of the matrix that is the outer parentheses on the left side.  

The resulting estimates of the raw, unsmoothed density and CDF of $x$ are $\hat{\cal R}(x)$ in Eq.~(\ref{eq9}) and 
\begin{equation}
\label{eq13}
\hat{R}(x)\equiv\sum\nolimits_{j=1}^N {w_j {\cal H}(x-x_j )}~~, \end{equation}
respectively. $\hat{P}(u)$ and ${\hat{P}}'(u)$ produce \textit{identical} results when evaluated at the sample points $\{u\}$. We recognize that they are not estimates but pure transformations of the sample, through the intermediary of the weights $\{w\}$. The same is true of $\hat{\cal R}$ and $\hat{R}$, which are sums of discontinuous functions. Indeed, the sum of delta functions in Eq.~(\ref{eq9}) is not particularly useful in itself because it conveys nothing more than the transformed sample.  We must smooth $\hat{\cal R}_(x)$ in order to calculate non-zero results for the density at values of the unprojected quantity other than the sample points.  We discuss this smoothing in the next section. 

\section{NONPARAMETRIC DENSITY ESTIMATION}

Here we explore the issues associated with smoothing the raw density, $\hat{\cal R}(x)$. Without smoothing, the form of $\hat{\cal R}(x)$ in Eq.~{(\ref{eq9}) is not very interesting or useful, because the only values it takes on are plus and minus infinity and zero. In order to pursue practical research with $\hat{\cal R}(x)$---such as comparing observations with theories, learning about possible variations of ${\cal R}(x)$ with other planetary or stellar parameters, and informing the measurement requirements for future missions and observing programs---we need $\hat{\cal R}(x)$ in the form of a sufficiently smoothed positive function. At the same time, we want to avoid over-smoothing, which might discard real detail. 

The astronomer's ``smoothing'' is called ``nonparametric density estimation'' in statistics and other fields (see Silverman 1986, Takezawa 2006, and Wasserman 2006). Our approach, of convolving the raw density with a Gaussian of standard deviation $\sigma$,
\begin{equation}
\label{eq14}
\hat{\cal R}_{\sigma}(x)=\sum\nolimits_{i=1}^m w_j \frac{1}{\sqrt{2\pi}\sigma} e^{-\frac{(x-u_i)^2}{2\sigma^2}}~~,
\end{equation}
is called ``kernel density estimation with a normal kernel of bandwidth 
$\sigma$.'' The standard statistical treatment flows from the study of histograms of samples of independent, identically distributed random variables, in the limit as (a)~the bin width tends to zero, (b)~the count in any bin becomes zero or one, (c)~the raw density is the sum of delta functions of equal, positive weight (1/$n$) located at the sample points, and (d)~the kernel density estimate is the sum of identical kernel functions located at the sample points. Ours is a different case, for which a theory must still be developed. Our raw density [Eq.~(\ref{eq9})] is the sum of unequally weighted delta functions, and the weights include both positive and negative values. These differences from the standard case occur because we are estimating density in the unprojected space---where the sample is represented by the unequal, positive and negative weights---rather than in the projected space of the sample itself.

A considerable literature has been developed on the tasks of selecting the kernel and bandwidth for the standard case. Optimizing the bandwidth usually calls for defining an objective function (figure of merit), which is maximized or minimized. Because the true density is not known, by definition, the objective function must be computed from the sample and bandwidth alone. It is not immediately clear how the extensive work on this problem in the standard case applies to the current case. Therefore, we take an empirical approach to bandwidth selection for a normal kernel by conducting Monte Carlo experiments with a theoretical true density. 

We expect the optimal value of $\sigma$ to be well-defined, because the asymptotes of Eq.~(\ref{eq14}) are trivial: reversion to Eq.~(\ref{eq9}) for  $\sigma\rightarrow0$ and approaching zero everywhere as $\sigma\rightarrow\infty$. Therefore, the optimal value of $\sigma$ must lie in between. In addition, we expect the optimal value to decrease with higher cardinality of the sample, $n$, because the increased information from more observations should include more information about detail. At least for $\sigma\ll\Delta u$, the metric for the range of $u$, we expect by simple scaling that the optimal value of $\sigma$ for different problems is proportional to $\Delta u$. Following Wasserman (2006; p.~135), we measure the range using the range metric in the Normal Reference Rule: $\Delta u\equiv \min(s,q/1.34)$, where $s$ is the sample standard deviation and $q$ is the interquartile range.

As a first exploration, we study the case of a theoretical density ${\cal R}(x)$ comprising three Gaussians, with $\{$mean, standard deviation, weight$\} = \{0.5, 0.25, 2.0\}$, $\{2.0, 0.5, 8.0\}$, and $\{2.5, 0.125, 1.0\}$. In this case, $\Delta u=0.783$. For this exercise we developed a facility to perform the following sequence of steps: (1)~create random samples $\{u\}$ of cardinality $n$, where each value is drawn from the random deviate or $u$; and for such samples, (2)~solve Eqs.~(\ref{eq12}) for weights $\{w\}$; (3)~construct the estimated density $\hat{\cal R}_{\sigma}(x)$ for any value of $\sigma$ using Eq.~(\ref{eq14}); (4)~compute the objective function (integrated square error) $v$ for the closeness of $\hat{\cal R}_{\sigma}(x)$ to ${\cal R}(x)$ , where
\begin{equation}
\label{eq15}
v\equiv\int_{-\infty}^{\infty} \left(\hat{\cal R}_{\sigma}(x)-{\cal R}(x)\right)^2 dx~~;
\end{equation}
and (5)~determine the value of $\sigma$ that minimizes $v$, which we adopt as the working definition of the optimal value of $\sigma$. 

The random deviate for $u$ is
\label{eq16}
\begin{equation}
\mathbb{U}=\log(10^\mathbb{X} \mathbb{Y})~~,
\end{equation}
where
\begin{equation}
\label{eq17}
\mathbb{X}=x\mathrm{-root}\left( \int_{-\infty}^{x} R(x')dx'=\mathbb{Z}\right)~~,
\end{equation}
where $R(x')$ stands for the true or estimated density under study, where
\begin{equation}
\label{eq18}
\mathbb{Y}=\sqrt{2\mathbb{Z}-\mathbb{Z}^2}~~,
\end{equation}
where $\mathbb{Z}$ is the uniform random deviate on the range 0--1, and where ``$x$-root'' in Eq.~(\ref{eq17}) is defined as the value of $x$ that satisfies the equation in parenthesis.

The results of experiments involving these five steps are shown in Figs.~1--6. Figure~\ref{fig1} confirms the expectation that lower values of $\sigma$ retain the spiky original pattern of delta functions in Eq.~(\ref{eq9}), and that higher values of $\sigma$ reduce contrast by blurring features. As suggested by the progression of the colored curves in Figure~\ref{fig1}, varying $\sigma$ to minimize $v$ will work efficiently to locate an optimal value, which for $n = 100$ should be somewhere in the range $0.1 < \sigma < 0.4$.
\begin{figure}
\centering
\plotone{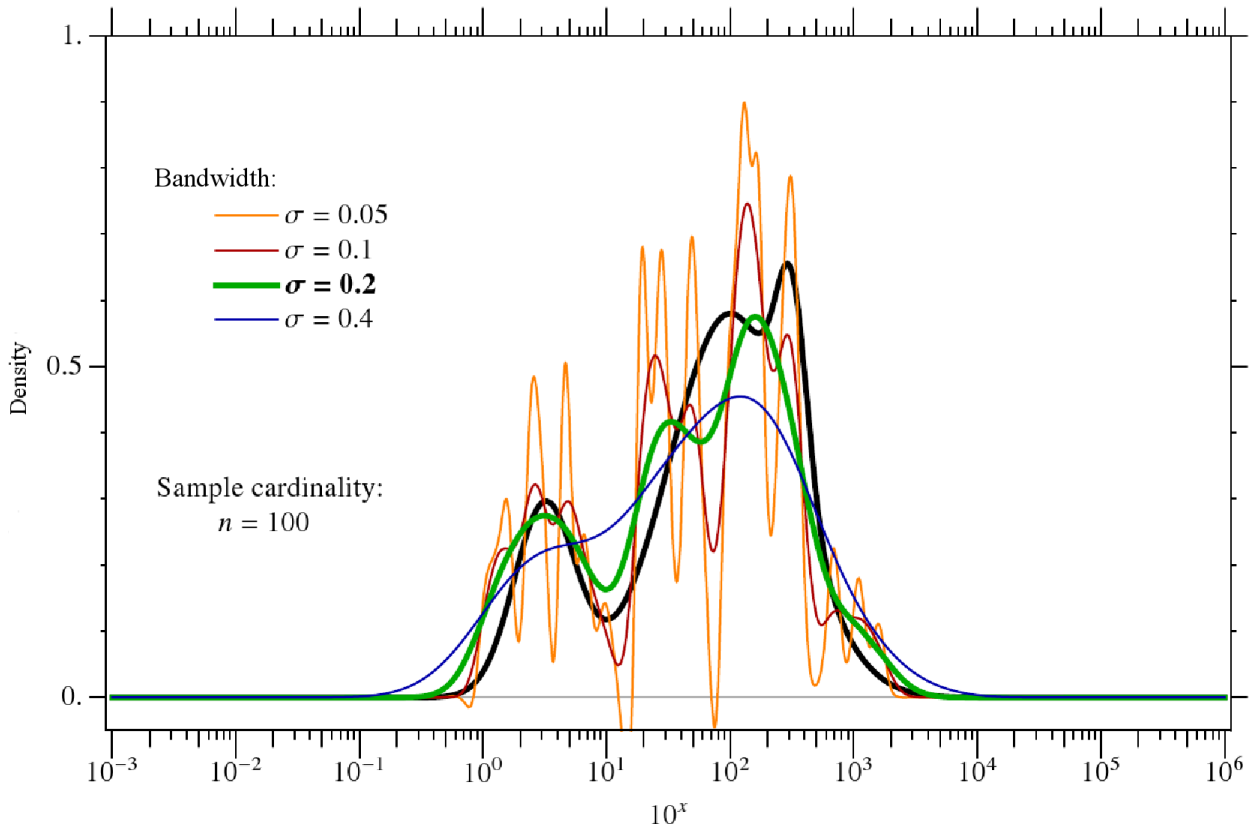}
\caption{For a theoretical three-Gaussian density ${\cal R}(x)$ (black curve), the effects of convolving the raw density $\hat{\cal R}(x)$ with normal kernels of various bandwidths $\sigma$. Four random samples $\{u\}$ with $n = 100$ were prepared using Eq.~(\ref{eq16}). The four sets of weights $\{w\}$ were determined by solving Eqs.~(\ref{eq12}). The four functions $\hat{\cal R}(x)$ resulting from Eq.~(\ref{eq9}) produced four functions $\hat{\cal R}_{\sigma}(x)$ from Eq.~(\ref{eq14}) with $\sigma = 0.05$, 0.10, 0.2, and 0.4, which are plotted here in color. The red and orange curves are under-smoothed and still dominated by vestiges of the delta functions. The blue curve is over-smoothed: the dip between the two main peaks is blurred away, and the tails extend well beyond the range of ${\cal R}(x)$. The green curve is the most satisfactory of the four, accurately locating the two main peaks in ${\cal R}(x)$ and revealing the distinct minimum between them. 
\label{fig1}}
\end{figure}
\begin{figure}
\centering
\plotone{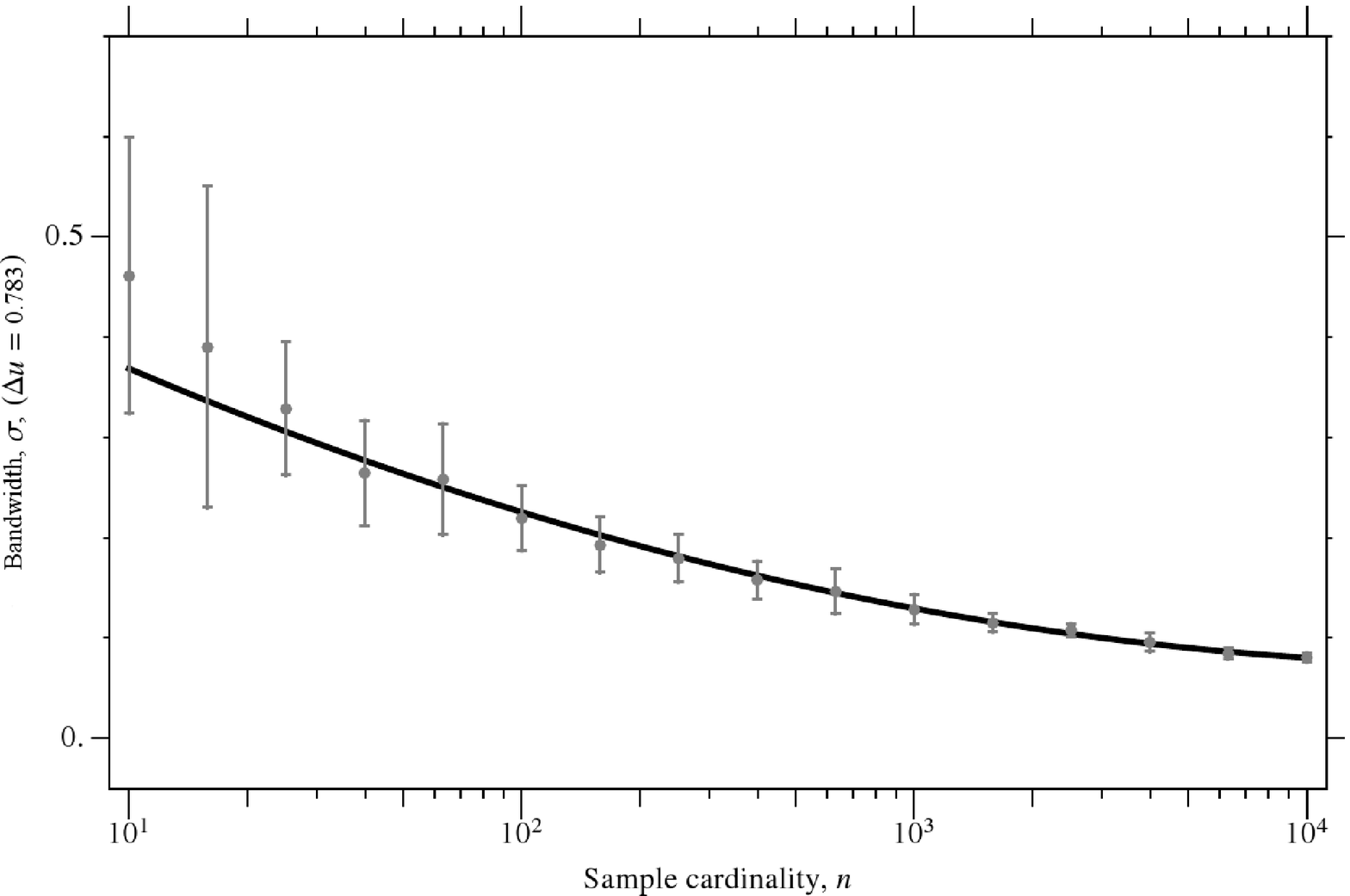}
\caption{Bandwidth calibration, $\sigma (n)$. Points: mean values of $\sigma$ that minimize the integrated square error, determined from 24 samples $\{u\}$ for each value of $n$. Error bars: the standard deviations of the 24 values contributing to each mean. Curve: the weighted quadratic fit to the data points. We use this calibration, scaled by the range metric, to select $\sigma$ for all smoothing in the remainder of this paper.
\label{fig2}}
\end{figure}

Figure~\ref{fig2} shows the results of minimizing $v$ to optimize $\sigma$. For each of 16 values of $\log n$ in the range 1--4, we prepared 24 random samples $\{u\}$ of cardinality $n$, drawing from the same three-Gaussian distribution. Next, we adjusted $\sigma$ to minimize $v$ in order to obtain the optimal value of $\sigma$ for each sample. Next, we determined the mean and standard deviation of the 24 values of optimized $\sigma$ at each value of $n$. Finally, we computed the best quadratic fit to these data points, which is
\begin{equation}
\label{eq19}
\sigma(n)=\left( 0.56-0.21\log n + 0.023(\log n)^2\right) \frac{\Delta u}{0.783}~~.
\end{equation}

We use Eq.~(\ref{eq19}) to compute the value of $\sigma$ in the remainder of this paper.

Figures~\ref{fig3}--\ref{fig6} illustrate the qualitative improvement in the resolution and accuracy of the amplitudes of $\hat{\cal R}_{\sigma(n)}(x)$ as $n$ increases. In this controlled experiment, with ${\cal R}(x)$ known, we can evaluate performance in two ways: \textit{absolutely}, by directly comparing $\hat{\cal R}_{\sigma(n)}(x)$ to the known ${\cal R}(x)$, and \textit{relatively}, by assessing the \textit{variation} of independent realizations of  $\hat{\cal R}_{\sigma(n)}(x)$ with respect to each other. In this experiment, these alternative evaluations are shown to be consistent: features that repeat robustly in multiple realizations are seen to correspond to true characteristics of ${\cal R}(x)$. Meanwhile, features that do not repeat are spurious. Said differently, we find no robustly repeated features in $\hat{\cal R}_{\sigma(n)}(x)$ that are not present in ${\cal R}(x)$, nor any evidence that $\hat{\cal R}_{\sigma(n)}(x)$ with sufficiently high $n$ would fail to reproduce---to any desired accuracy---any feature in ${\cal R}(x)$, no matter how narrow or small in amplitude.

\begin{figure}
\centering
\plotone{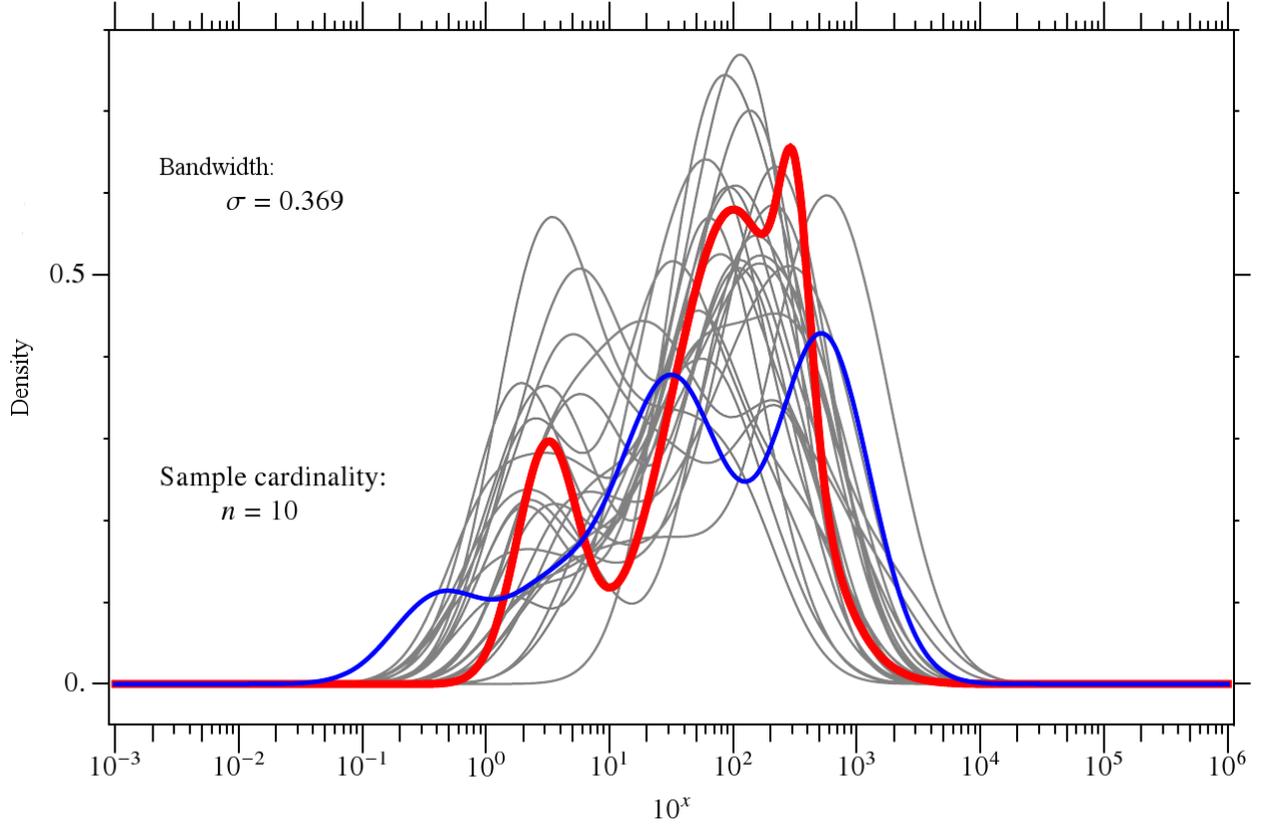}
\caption{The variation of $\hat{\cal R}_{\sigma}(x)$ for $n = 10$, in the case of the same theoretical three-Gaussian distribution ${\cal R}(x)$ (red curve). Gray and blue curves: 25 independent, random realizations of $\hat{\cal R}_{\sigma (n)}(x)$. There are no robustly repeated features in the recovered density that are present in the true density, except the extent and perhaps the skewness of  ${\cal R}(x)$. This result suggests that RV or ML samples with $n = 10$ over a comparable range should offer little information about the true distribution of mass or orbital radius. The fractional amplitude noise at the peak is about $\pm$40\%.
\label{fig3}}
\end{figure}

In Figure~\ref{fig3}, we find that samples with $n = 10$ offer little information about the true distribution of $x$, with fluctuations on the scale of $\sim$40\% amplitude near the peak. In Figure~\ref{fig4}, we find that samples with $n = 100$ offer crude information about the shape of the true distribution, with fluctuations on the scale of $\sim$20\% amplitude near the peak. In Figure~\ref{fig5}, we find that samples with $n = 1000$ reveal the basic structure of the true distribution, with fluctuations on the scale of $\sim$10\% amplitude near the peak. In Figure~\ref{fig6}, resolution has improved, noise has been reduced, and some substructure is revealed, with fluctuations on the scale of $\sim$5\% amplitude near the peak.
\begin{figure}
\centering
\plotone{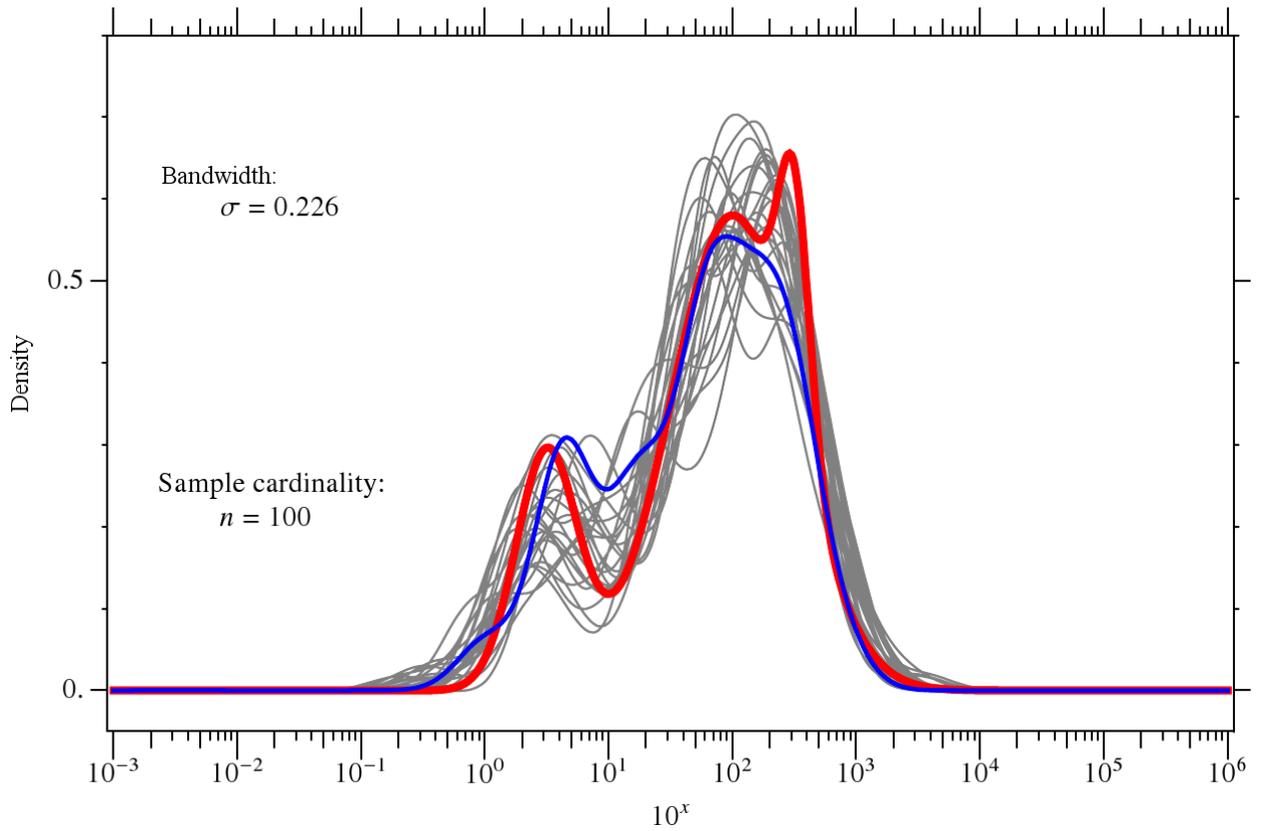}
\caption{The variation of  $\hat{\cal R}_{\sigma}(x)$ for $n = 100$. Amplitude variation is reduced and resolution is increased. The bimodal character of ${\cal R}(x)$ is revealed. We see that RV or ML samples with $n = 100$ should contain crude information about the shape of the true distribution of mass or orbital radius. The fractional amplitude noise at the peak is about $\pm$20\%.
\label{fig4}}
\end{figure}
\begin{figure}
\centering
\plotone{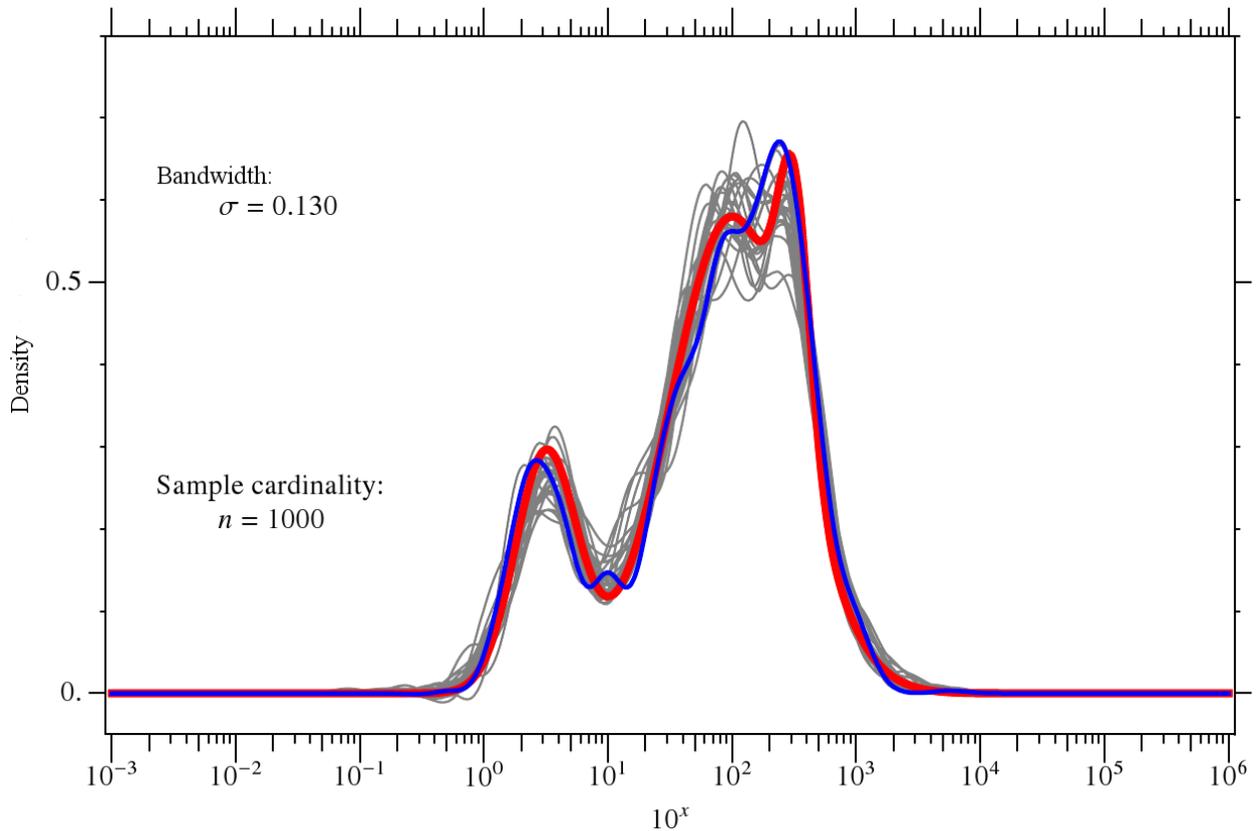}
\caption{The variation of $\hat{\cal R}_{\sigma}(x)$ for $n = 1000$. The locations and amplitudes of the two main peaks are accurately recovered. Improved resolution reveals the greater width of the right-hand peak compared with the left-hand peak, but the dip at the top is not recovered. Amplitude variation is further reduced. We find that RV or ML samples with $n = 1000$ should convey information about the basic structure of the true distribution of mass or orbital radius, but would not reveal secondary features with less than about 10\% amplitude, which is approximately the fractional noise amplitude at the peak.
\label{fig5}}
\end{figure}
\begin{figure}
\centering
\plotone{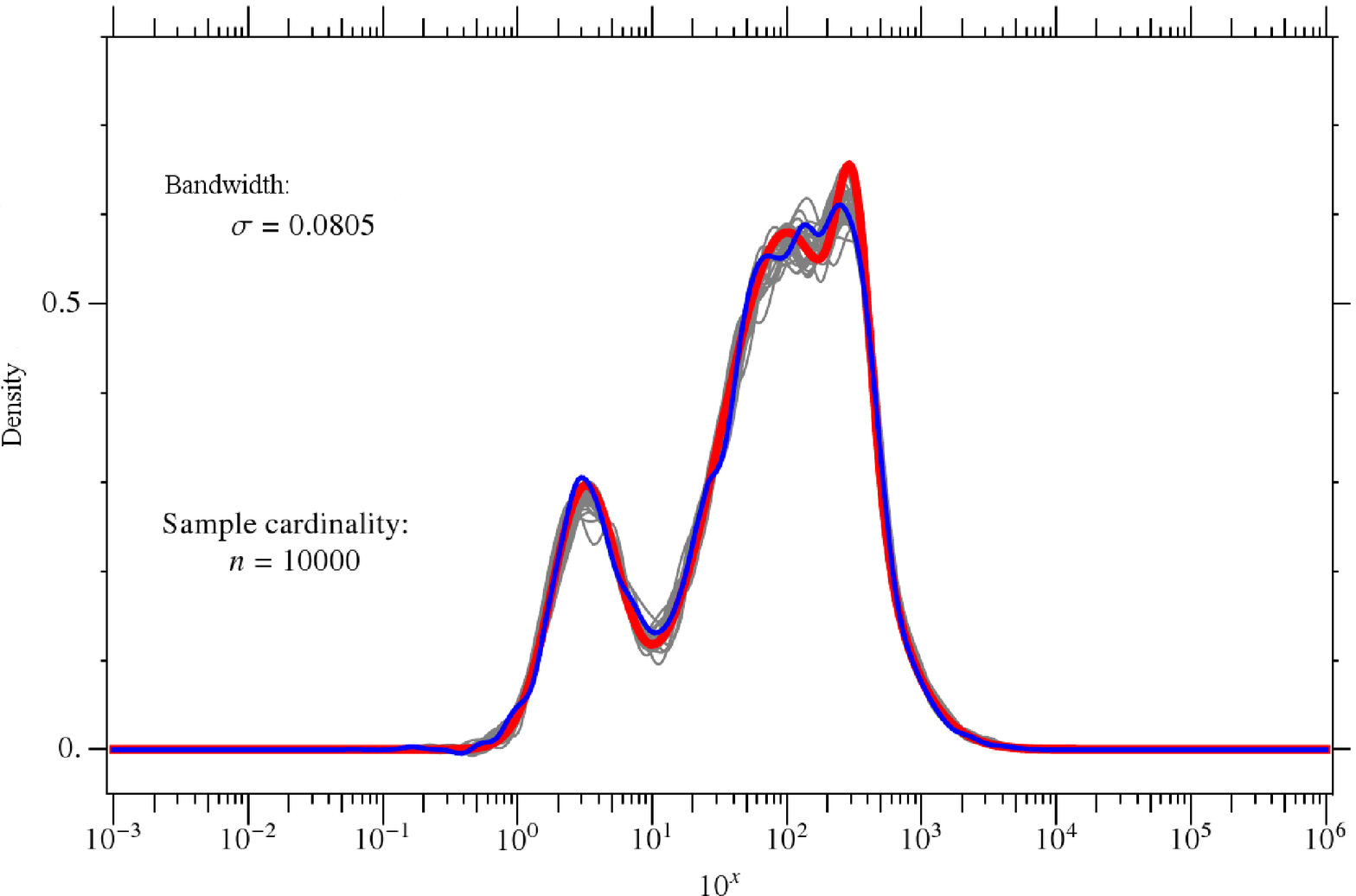}
\caption{The variation of  $\hat{\cal R}_{\sigma}(x)$ for $n = 10,000$. The amplitude accuracy and resolution of individual realizations are further improved, and the small dip at the top of the right peak is marginally resolved. The fraction amplitude of noise at the peak is about $\pm$5\%.
\label{fig6}}
\end{figure}

We can summarize these experiments with the finding that the fractional noise fluctuations at the peak, expressed as a percentage of the peak height, are approximately 80~$n^{-\log 2}$.

In the following section, we study real data from RV programs, where the true function ${\cal R}(x)$ is unknown---and indeed is the main object of the research. We require a method to determine which features are believable in the density $\hat{\cal R}_{\sigma(n)}(x)$ estimated from such data. For this we need the functional equivalent of a density of densities, to describe the distribution of $\hat{\cal R}_{\sigma(n)}(x)$, which is the distribution we might estimate from multiple, statistically equivalent data sets drawn from nature---but which we do not have, of course. The way forward is to assume that the unobtainable distribution is approximately the same as the distribution of functions $\hat{\cal R}_{\sigma(n)}(x)$ that we \textit{can} compute from multiple samples \textit{drawn from the random deviate for the density} $\hat{\cal R}_{\sigma(n)}(x)$ \textit{derived from the real data}. 

Basically, this approach uses self-consistency as a measure of the fidelity of the inferred density. It assumes that $\hat{\cal R}_{\sigma(n)}(x)$ is the true density and asks whether new, random data sets drawn from $\hat{\cal R}_{\sigma(n)}(x)$, with the same cardinality, robustly repeat the features in $\hat{\cal R}_{\sigma(n)}(x)$. If so, they are probably real. If not, they are probably spurious.

This approach to ``confidence'' in $\hat{\cal R}_{\sigma(n)}(x)$ is a natural extension of the Monte Carlo methods for determining confidence regions described in Section~14.5 of Press et~al.\ (1986). We take this approach in the next section, where we study two real samples of RV data.

\section{RV DATA}

To illustrate the method developed in Sections~3--4, we treat two samples of RV data. The first is the 67 values of $m \sin i$ treated by Jorissen et~al.\  (2001), which were kindly provided to us by A.~Jorissen. This sample is of particular interest because the authors used it to estimate the density for $m$ using Eq.~(\ref{eq5}) (which is identical to their Eq.~4 if completeness is ignored [$C(u)\equiv 1$] and the logarithmic variables are changed back to $m$ and $m \sin i$). Therefore, we can compare results. 

Figure~\ref{fig7} shows our result for $\hat{\cal R}_{\sigma (n)}(\log m$), which is nearly featureless, unlike the Jorissen et~al.\ results, shown in their Fig.~2. Because they do not use logarithmic variables, their results are simplified for $m<1.\,m_\mathrm{J}$, even though 31\% of the sample values lie in that range. Nevertheless, we can recover the main features in the Jorissen et~al.\ PDF for  $m>1.\,m_\mathrm{J}$ \textit{if we grossly under-smooth}. 
\begin{figure}
\centering
\plotone{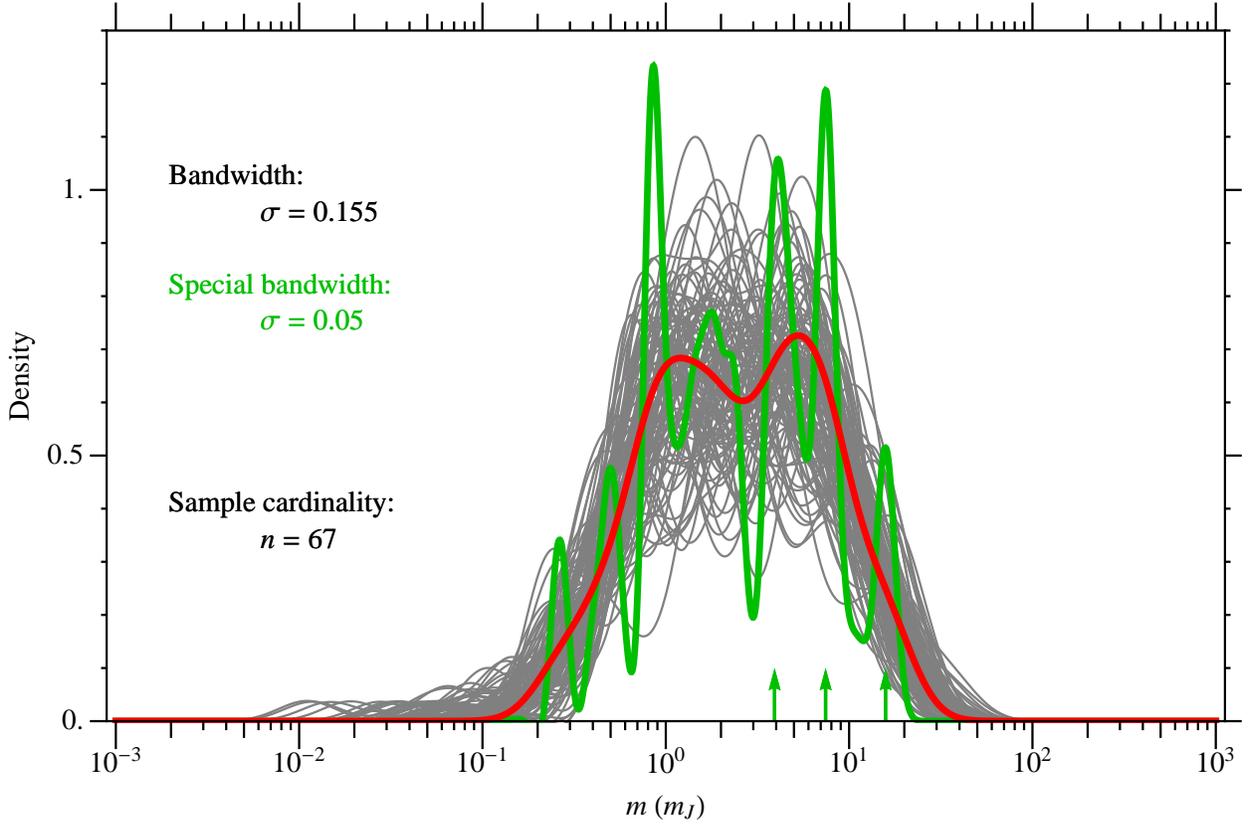}
\caption{Treatment of the 67 RV data points of Jorissen et~al.\ (2001). The value $\sigma = 0.155$ comes from Eq.~(\ref{eq19}) using $\Delta u = 0.491$, the range metric of the data. In red, the inferred metric, $\hat{\cal R}_{\sigma (n)}(\log m$). In gray, the confidence region delineated by densities generated from 100 statistically equivalent data sets drawn from the random deviate for  $\hat{\cal R}_{\sigma (n)}(\log m$). The dip in the peak of the red curve is not believable, given the significantly greater width of the confidence region (gray). In green, an under-smoothed version of  $\hat{\cal R}_{\sigma}(\log m$), with $\sigma = 0.05$, which resembles the density inferred by Jorissen et~al.\ (2001), for $m > 1\, m_\mathrm{J}$, shown in their Fig.~2, particularly the peaks at $\sim$4.0, 7.5, and 15\,$m_\mathrm{J}$, indicated here by green arrows. These features are apparently artifacts of under-smoothing.
\label{fig7}}
\end{figure}

The feature Jorissen et~al.\ found most believable is the minimum between the second and third peaks (green arrows), in the range 10--13\,$m_\mathrm{J}$. They applied a ``jackknife'' test and concluded that this minimum was a ``robust result, not affected by the uncertainty in the solution.'' Based on our experiments in Section~4, and on our treatment of the Jorissen data (red and gray in Fig.~\ref{fig7}), all the features in our under-smoothed  $\hat{\cal R}_{\sigma (n)}(\log m$), including the minimum favored by Jorissen et~al., which we can reproduce, are artifacts of under-smoothing and spurious.

The dip in the peak of the red curve in Figure~\ref{fig7} is not believable, given the significantly greater width of the confidence region (gray). 

The second sample we studied is the 308 values of $m \sin i$ listed at exoplanets.org on 20 October 2010. This sample is documented in Table~1, and Figure~\ref{fig8} shows the results of our treatment. The feature favored by Jorissen et~al.\ at 10--13\,$m_\mathrm{J}$ is also not present in this density, which is improved in both resolution and amplitude accuracy compared with $n = 67$.

\begin{figure}
\centering
\plotone{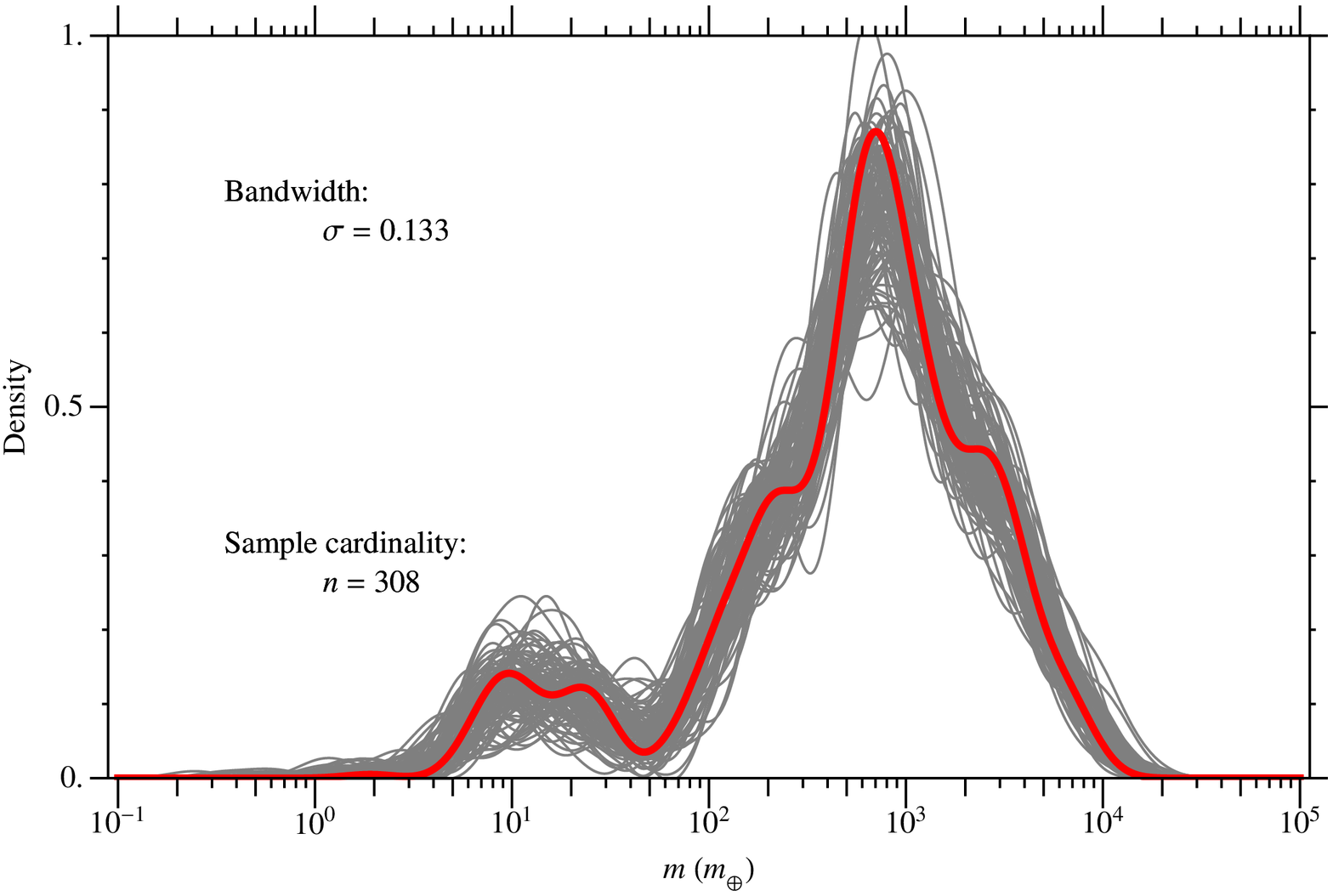}
\caption{Treatment of the sample of 308 values of $m \sin i$ from RV observations listed at exoplanets.org on 20 October 2010. The sample is documented in Table~1. The value $\sigma = 0.133$ comes from Eq.~(\ref{eq19}) using $\Delta u = 0.600$, the approximate range of the data. In red, the inferred density, $\hat{\cal R}_{\sigma (n)}(\log m$). In gray, the confidence region delineated by densities generated from 100 statistically equivalent data sets, which were drawn from the random deviate for $\hat{\cal R}_{\sigma (n)}(\log m$). The rise of the density with decreasing $m$ from $\sim$50--10\,$m_{\oplus}$ (0.16--0.03\,$m_\mathrm{J}$) repeats robustly and is statistically significant.
\label{fig8}}
\end{figure}

For the $n = 308$ density, the increase as $m$ decreases from $\sim$50 to $\sim$10\,$m_{\oplus}$ (0.16--0.03\,$m_\mathrm{J}$) repeats robustly in Monte Carlo trials and is statistically significant. Interestingly, this minimum is near the low-$m$ cutoff of the Jorissen data, and therefore no information about the bimodal character of  $\hat{\cal R}_{\sigma (n)}(x)$ is present in the earlier sample of $n = 67$ RV data points.

\section{COMMENTS AND FUTURE DIRECTIONS}

This paper offers an improved understanding of statistical inferences regarding the density of the unprojected quantities---mass $m$ for RV or radial distance $r$ for ML---from samples of the projected quantities, $m \sin i$ or $r \sin\beta$. The same treatment applies to both ML and RV. 

We find that that ability to confidently recognize real features in the estimated density---and to reject spurious ones---depends on (1)~the scale of the variations compared with the blurring scale of ${\cal Q}$ (standard deviation of 0.18 in the logarithm), (2)~$\Delta u$, the metric of the range of the measurements, (3)~the amplitudes of the features, and (4)~$n$, the cardinality of the sample. The latter is the only factor controlled by the observer or the design of a mission. 

Care must be given to the completeness factor, ${\cal C}(u)$, as well as to other possible systematic effects and biases that might affect the fidelity with which a sample of projected values reflects the true distribution of the unprojected quantity. For example, in the case of RV, we know that the detection efficiency---and therefore the completeness---decreases to zero as the projected mass---and therefore the signal-to-noise ratio ($u$ divided by the noise amplitude) decreases. Also, the least-squares estimator for mass from astrometry is known to be biased at low signal-to-noise ratio, and the same can be expected for $m \sin i$ from RV, as the treatment of all Keplerian signals is basically the same for planet detection by periodogram. (Brown 2009; see also Hogg et~al.\ 2010) Any such factors affecting ML samples also bear close scrutiny. Monte Carlo experiments with the random deviates and analytic method developed in this paper should be helpful for studying the systematic effects of ${\cal C}(u)$ on the density.

Astro 2010 recommended an unbiased census of Earth-like exoplanets by \textit{WFIRST} using the ML technique. This recommendation implies specific---but not yet defined---science requirements on the accuracy of the $\hat{\cal R}_m (r)$ inferred from ML events by exoplanets with typical characteristics $m\approx1\,m_{\oplus}$ and $r\approx1$~AU. In turn, these science requirements will imply measurement requirements---particularly on $n$, which controls random errors, and on knowledge of ${\cal C}(u)$, which controls systematic errors. In turn, the measurement requirements will flow down to the mission design, including both spacecraft and ground operations. We expect that Monte Carlo studies based on the method in this paper will be helpful in achieving adequacy and self consistency for the ML components of the \textit{WFIRST} project, from spacecraft to ground operations.

We note that in the usual case where the angular factor $y$ is not known independently, Ho and Turner (2011) have recently pointed out that one must assume a density for the unprojected quantity $x$ in order to properly state the confidence interval for the value of $x$ derived from any particular observation of the projected quantity $u$. This is because the posterior distribution of $y$ is not the same as the prior distribution of $y$. The needed density could come from (1)~a theoretical guesstimate (entailing systematic uncertainties), (2)~a sample of $x$ from planets with known $y$ (transiting planets in the case of RV), or (3)~a sample of $u$ from planets with unknown $y$, using the method developed in this paper.

The introduction of the logarithmic variables $u$, $x$, and $\log y$ in Section~2, 
\begin{equation}
u=x+\log y~~,
\end{equation}
changed the problem from one of a product of random variables to one of a sum of random variables. This change creates a connection to recent statistical research on the ``errors-in-variables'' problem (Studenmayer et~al.\ 2008; Apanasovich et~al.\ 2009; Delaigle et~al.\ 2009). The goal in this problem is to estimate the density of $x$, which is not observable, from observations of $u$, which is a version of $x$ contaminated by the additive, homoscedastic measurement error, $\log y$, with known density. The cited papers explore non-parametric estimation of the density of $x$ variously using B~splines, simulation extrapolation, and local-polynomials. We expect that future research into density estimation for $m$ or $r$ from RV and ML samples of $m \sin i$ or $r \sin\beta$ will explore such advances in statistical research, and produce instructive comparisons with our approach---kernel density estimation with a normal kernel. 

\acknowledgements{We thank D.~Latham, D.~Spiegel, W.~Traub, E.~Turner, and an anonymous referee for their helpful comments. We thank A.~Jorissen for providing the sample of 67 RV measurements of $m \sin i$ used in Jorissen et~al.\ (2001). We thank Sharon Toolan for her expert preparation of the manuscript.}

\renewcommand\baselinestretch{1}
\begin{deluxetable}{lcclcclc}
\tablecaption{The 308 Values of $m \sin i$ from RV at exoplanets.org on 20 October 2010}
\tabletypesize{\small}
\tablecolumns{6}
\tablehead{
\colhead{Exoplanet} &\colhead{$m\sin i (R_\oplus)$} &\colhead{~}
&\colhead{Exoplanet} &\colhead{$m\sin i (R_\oplus)$} &\colhead{~}
&\colhead{Exoplanet} &\colhead{$m\sin i (R_\oplus)$} 
}   
\startdata
GJ 581 e &\phn1.943 &&HD 45364 b &\phn59.490 &&47 UMa c &173.489\\
HD 40307 b &\phn4.103 &&HD 107148 b &\phn67.444 &&HD 175541 b &182.742\\
HD 156668 b &\phn4.151 &&HD 46375 b &\phn72.222 &&HIP 14810 d &184.560\\
61 Vir b &\phn5.110 &&HD 3651 b &\phn72.786 &&HD 27894 b &196.424\\
GJ 581 c &\phn5.369 &&HD 76700 b &\phn73.793 &&HD 11964 b &196.508\\
GJ 876 d &\phn5.893 &&HD 168746 b &\phn77.913 &&GJ 876 c &196.761\\
HD 215497 b &\phn6.629 &&HD 16141 b &\phn79.359 &&HD 330075 b &198.268\\
HD 40307 c &\phn6.722 &&HD 108147 b &\phn82.012 &&HD 37124 d &202.306\\
GJ 581 d &\phn7.072 &&HD 109749 b &\phn87.311 &&HD 192263 b &203.245\\
HD 181433 b &\phn7.544 &&HIP 57050 b &\phn94.643 &&HD 181433 c &203.574\\
HD 1461 b &\phn7.630 &&HD 101930 b &\phn95.105 &&GJ 832 b &204.802\\
55 Cnc e &\phn7.732 &&HD 88133 b &\phn95.187 &&HD 216770 b &205.644\\
GJ 176 b &\phn8.264 &&GJ 649 b &103.419 &&HD 37124 b &206.867\\
HD 40307 d &\phn9.100 &&HD 215497 c &104.113 &&HD 45364 c &209.326\\
HD 7924 b &\phn9.256 &&BD $-$08 2823 c &104.296 &&HD 170469 b &212.557\\
HD 69830 b &10.060 &&HD 33283 b &104.954 &&$\upsilon$ And b &212.747\\
61 Vir c &10.609 &&HD 47186 c &110.712 &&HD 96167 b &217.609\\
$\mu$ Ara d &10.995 &&HD 164922 b &113.777 &&HD 209458 b &218.862\\
GJ 674 b &11.087 &&HD 149026 b &114.129 &&HD 37124 c &221.403\\
HD 69830 c &11.689 &&HD 93083 b &117.035 &&HD 9446 b &222.123\\
BD $-$08 2823 b &14.604 &&HD 181720 b &118.247 &&HD 224693 b &227.153\\ 
HD 4308 b &15.175 &&HD 63454 b &122.451 &&HD 34445 b &239.582\\
GJ 581 b &15.657 &&HD 126614 A b &122.607 &&HD 187085 b &255.476\\
HD 69830 d &17.908 &&HD 83443 b &125.835 &&HD 134987 c &255.868\\
HD 190360 c &18.746 &&HD 212301 b &125.927 &&HD 4208 b &256.672\\
HD 219828 b &19.778 &&HD 6434 b &126.259 &&GJ 179 b &262.008\\
HD 16417 b &21.285 &&BD $-$10 3166 b &136.656 &&GJ 849 b &263.683\\
HD 47186 b &22.634 &&HD 102195 b &143.935 &&55 Cnc b &268.236\\
61 Vir d &22.758 &&HD 75289 b &146.236 &&HD 38529 b &272.456\\
GJ 436 b &23.490 &&51 Peg b &146.587 &&HD 155358 b &284.379\\
HD 11964 c &24.905 &&HD 45652 b &148.909 &&HD 179949 b &286.749\\
HD 179079 b &26.631 &&HD 2638 b &151.731 &&HD 10647 b &294.048\\
HD 49674 b &32.287 &&HD 155358 c &160.167 &&HD 114729 b &300.317\\
HD 99492 b &33.752 &&HD 99109 b &160.273 &&HD 185269 b &303.322\\
55 Cnc f &46.285 &&HD 187123 b &162.135 &&HD 154345 b &304.180\\
HD 102117 b &53.962 &&HD 208487 b &162.817 &&HD 108874 c &326.896\\
55 Cnc c &54.254 &&HD 181433 d &170.218 &&HD 60532 b &328.940\\
HD 117618 b &56.154 &&$\mu$ Ara e &172.710 &&HD 130322 b &331.575\\%
$\rho$ CrB b &338.250 &&16 Cyg B b &521.310 &&HD 147018 b &\phn676.160\\
HD 52265 b &340.571 &&HD 167042 b &522.435 &&HD 118203 b &\phn679.059\\
$\epsilon$ Eri b &344.932 &&HD 4113 b &523.954 &&HD 216437 b &\phn689.212\\
HD 231701 b &345.428 &&HD 142415 b &528.293 &&HD 206610 b &\phn707.519\\
HD 114783 b &351.270 &&HD 82943 b &551.047 &&HD 212771 b &\phn715.885\\
HD 189733 b &362.504 &&HD 50499 b &554.610 &&HD 202206 c &\phn740.966\\
HD 73534 b &367.323 &&$\mu$ Ara b &554.861 &&HD 154857 b &\phn741.503\\
HD 100777 b &370.334 &&HD 87883 b &558.101 &&HD 12661 b &\phn744.088\\
GJ 317 b &373.522 &&HD 20782 b &560.637 &&HD 4313 b &\phn746.348\\
HD 147513 b &374.984 &&$\gamma$ Cep b &563.066 &&HD 192699 b &\phn758.958\\
HD 48265 b &383.503 &&HD 89307 b &563.295 &&HD 73526 c &\phn769.550\\
HD 148427 b &385.455 &&HD 68988 b &572.154 &&HD 23079 b &\phn776.670\\
HD 216435 b &386.126 &&HD 74156 b &573.902 &&HD 60532 c &\phn782.970\\
HD 65216 b &386.653 &&HD 8574 b &574.096 &&HD 43691 b &\phn793.794\\
HD 121504 b &388.542 &&HD 9446 c &577.1001 &&HD 75898 b &\phn799.475\\
HD 95089 b &392.849 &&HD 117207 b &578.187 &&47 UMa b &\phn809.281\\
HD 210277 b &404.579 &&HD 171028 b &581.417 &&HD 41004 A b &\phn812.705\\
HIP 14810 c &405.351 &&HD 45350 b &583.668 &&HD 171238 b &\phn829.345\\
HD 108874 b &410.126 &&HD 73256 b &594.119 &&HD 217107 c &\phn831.396\\
HD 142 b &415.053 &&HD 13931 b &598.001 &&HD 62509 b &\phn854.159\\
HD 19994 b &421.763 &&$\mu$ Ara c &600.493 &&HD 164604 b &\phn854.614\\
HD 149143 b &422.815 &&HD 190647 b &604.908 &&HD 81688 b &\phn855.313\\
HD 30562 b &423.584 &&HD 70642 b &606.968 &&11 Com b &\phn864.016\\
HD 114386 b &433.490 &&$\upsilon$ And c &609.864 &&HD 153950 b &\phn871.649\\
HD 217107 b &445.384 &&GJ 876 b &614.349 &&HD 66428 b &\phn874.042\\
HD 23127 b &446.577 &&HD 5319 b &615.961 &&$\xi$ Aql b &\phn892.158\\
HIP 5158 b &453.398 &&HD 187123 c &617.382 &&HD 196050 b &\phn903.853\\
HD 128311 b &463.211 &&HD 12661 c &619.448 &&HD 73526 b &\phn907.731\\
HD 188015 b &467.143 &&HD 210702 b &624.623 &&HD 37605 b &\phn908.618\\
HD 205739 b &472.709 &&HD 5388 b &624.658 &&HD 169830 b &\phn918.494\\
HD 86081 b &475.529 &&$\kappa$ CrB b &629.334 &&HD 196885 b &\phn935.799\\
BD +14 4559 b &482.948 &&HD 82943 c &632.344 &&HD 181342 b &\phn953.480\\
HD 177830 b &487.061 &&HD 136418 b &633.689 &&HD 143361 b &\phn964.755\\
HD 190360 b &487.980 &&HD 20868 b &638.619 &&HD 73267 b &\phn973.522\\
$\epsilon$ Ret b &491.944 &&$\iota$ Hor b &650.641 &&HD 221287 b &\phn990.300\\
HD 180902 b &494.977 &&6 Lyn b &657.736 &&HD 72659 b &1003.123\\
HD 134987 b &496.992 &&HD 4203 b &661.869 &&HD 17156 b &1024.490\\
HD 159868 b &516.287 &&HIP 79431 b &671.672 &&HD 128311 c &1032.619\\%
HD 32518 b &1063.456 &&14 Her b &1651.132 &&$\gamma$ Leo A b &2802.989\\
HD 1237 b &1072.773 &&81 Cet b &1697.742 &&$\iota$ Dra b &2803.565\\
HD 183263 c &1105.088 &&HD 132406 b &1781.687 &&HD 33564 b &2901.538\\
HD 125612 b &1119.434 &&HD 145377 b &1837.971 &&HD 33636 b &2946.568\\
HD 92788 b &1132.890 &&HD 28185 b &1842.802 &&HD 141937 b &3011.953\\
HD 183263 b &1136.237 &&HD 102272 b &1879.590 &&HD 30177 b &3079.540\\
HD 195019 b &1137.966 &&HD 2039 b &1883.421 &&30 Ari B b &3139.968\\
HD 190984 b &1190.820 &&HD 190228 b &1888.806 &&HD 139357 b &3202.743\\
HIP 14810 b &1231.588 &&HD 10697 b &1981.982 &&HD 39091 b &3206.748\\
HD 80606 b &1236.588 &&HD 11977 b &2072.545 &&18 Del b &3245.402\\
42 Dra b &1237.297 &&HD 104985 b &2082.307 &&HD 38801 b &3420.489\\
55 Cnc d &1261.592 &&HD 147018 c &2095.941 &&HD 156846 b &3499.068\\
GJ 86 b &1271.840 &&HD 86264 b &2106.695 &&11 UMi b &3524.403\\
HD 40979 b &1278.579 &&HD 81040 b &2185.897 &&HD 136118 b &3713.095\\
HD 169830 c &1291.466 &&HD 111232 b &2200.074 &&HD 114762 b &3713.985\\
HD 204313 b &1291.660 &&HD 106252 b &2212.151 &&HD 38529 c &4157.806\\
$\upsilon$ And d &1308.326 &&4 UMa b &2267.236 &&BD +20 2457 c &4187.622\\
$\tau$ Boo b &1309.541 &&HD 240210 b &2316.907 &&HD 16760 b &4577.149\\
HD 16175 b &1392.132 &&HD 178911 B b &2317.730 &&HD 162020 b &4836.124\\
HD 50554 b &1398.267 &&70 Vir b &2371.824 &&HD 202206 b &5348.098\\
HD 142022 b &1420.156 &&$\epsilon$ Tau b &2422.334 &&HD 168443 c &5572.360\\
HD 213240 b &1440.776 &&HD 222582 b &2425.415 &&HD 131664 b &5826.122\\
14 And b &1488.884 &&HD 23596 b &2461.236 &&HD 41004 B b &5853.237\\
HD 11506 b &1505.054 &&HD 168443 b &2465.680 &&HD 137510 b &6936.118\\
HD 17092 b &1577.327 &&HD 175167 b &2472.565 &&BD +20 2457 b &7207.109\\
HD 154672 b &1591.415 &&HD 74156 c &2572.856 &&HD 43848 b &7729.763\\
HIP 2247 b &1628.547 &&HD 89744 b &2693.210\\
\enddata
\end{deluxetable}


\begin{references}

\reference{}Apanasovich, T.~V., Carroll, R.~J., \& Maity, A. 2009, ``SIMEX and standard error estimation in semiparameteric measurement error,'' Electronic~J.\ Statist., 3, 318--348


\reference{}Blandford, R.~D., \& Committee for a Decadal Survey of Astronomy and Astrophysics 2010, ``New Worlds, New Horizons in Astronomy and Astrophysics,'' (National Academies Press: Washington, D.C.)

\reference{}Brown, J.~R., \& Harvey, M.~E. 2007, Journal of Statistical Software, 19, 1 

\reference{}Brown, R.~A. 2009, ApJ, 699, 711

\reference{}Chandrasekhar, S., \& M\"unch, G. 1950, ApJ, 111, 142

\reference{}Dempster, A.~P., Laird, N.~M., \& Rubin, D.~B. 1977, ``Maximum Likelihood from Incomplete Data via the EM Algorithm,'' Journal of the Royal Statistical Society, Series B (Methodological) 39(1), 1--38. JSTOR 2984875. MR0501537

\reference{}Delaigle, A., Fan, J., \& Carroll, R.~J. 2009, ``A design-adaptive local polynomial estimator for the errors-in-variables problem,'' J.~Amer.\ Stat.\  Assoc., 104, 348--359

\reference{}Durban, J. 1968, The Annals of Mathematical Statistics, 39, 398

\reference{}Gaudi, B.~S. 2011, ``Microlensing by Exoplanets,'' in Exoplanets, S.~Seager, ed., (Tuscon: University of Arizona Press), pp. 79--110

\reference{}Glen, A.~G., Leemis, L.~M., \& Drew, J.~H. 2004, Computational Statistics and Data Analysis, 44, 451

\reference{}Ho, S., \& Turner, E.~L. 2011, submitted to ApJ

\reference{}Hogg, D.~W., Myers, A.~D., \& Bovy, J. 2010, arXiv:1008.4146v2

\reference{}Jorissen, A., Mayor, M., \& Udry, S. 2001, A\&A, 379, 992

\reference{}Press, W.~H., Flannery, B.~P., Teukolksky, S.~A., \& Vetterling, W.~T. 1986, Numerical Recipes (New York: Cambridge Univ.\ Press)

\reference{}Silverman, B.~W. 1986, Density Estimation for Statistics and Data Analysis (New York: Chapman and Hall)

\reference{}Staudenmayer, J., Ruppert, D., \& Buonaccorsi, J.~P. 2008, ``Density estimation in the presence of heteroscedastic measurement error,'' J.~Amer.\ Stat.\ Assoc., 103, 726--736

\reference{}Takezawa, K. 2006, Introduction to Nonparametric Regression (Hoboken: John Wiley \& Sons)

\reference{}Wasserman, L. 2006, All of Nonparametric Statistics (New York: Springer Science + Business Media)

\end{references}
\end{document}